\documentclass[english,onecolumn,pre,aps,showpacs]{revtex4}
\usepackage[T1]{fontenc}
\usepackage[latin1]{inputenc}
\usepackage{graphicx}

\makeatletter
\@ifundefined{textcolor}{}
{%
 \definecolor{BLACK}{gray}{0}
 \definecolor{WHITE}{gray}{1}
 \definecolor{RED}{rgb}{1,0,0}
 \definecolor{GREEN}{rgb}{0,1,0}
 \definecolor{BLUE}{rgb}{0,0,1}
 \definecolor{CYAN}{cmyk}{1,0,0,0}
 \definecolor{MAGENTA}{cmyk}{0,1,0,0}
 \definecolor{YELLOW}{cmyk}{0,0,1,0}
 }

\def\vec#1{\mbox{\boldmath $\mathrm{#1}$}}
\def\Pm{\mathrm{Pm}}
\def\Rm{\mathrm{Rm}}
\def\RE{\mathrm{Re}}
\def\Ro{\mathrm{Ro}}

\def\Re{\mathrm{Re}}
\def\Im{\mathrm{Im}}

\makeatother

\usepackage{babel}

\begin{document}

\title{Inviscid helical magnetorotational instability in cylindrical Taylor-Couette
flow}

\author{J\={a}nis Priede }

\affiliation{Applied Mathematics Research Centre, Coventry University, Coventry,
CV1 5FB, United Kingdom}

\email{J.Priede@coventry.ac.uk}
\begin{abstract}
This paper presents the analysis of axisymmetric helical magnetorotational
instability (HMRI) in the inviscid limit, which is relevant for astrophysical
conditions. The inductionless approximation defined by zero magnetic
Prandtl number is adopted to distinguish the HMRI from the standard
MRI in the cylindrical Taylor-Couette flow subject to a helical magnetic
field. Using a Chebyshev collocation method convective and absolute
instability thresholds are computed in terms of the Elsasser number
for a fixed ratio of inner and outer radii $\lambda=2$ and various
ratios of rotation rates and helicities of the magnetic field. It
is found that the extension of self-sustained HMRI modes beyond the
Rayleigh limit does not reach the astrophysically relevant Keplerian
rotation profile not only in the narrow- but also in the finite-gap
approximation. The Keppler limit can be attained only by the convective
HMRI mode provided that the boundaries are perfectly conducting. However,
this mode requires not only a permanent external excitation to be
observable but also has a long axial wave length, which is not compatible
with limited thickness of astrophysical accretion disks. 
\end{abstract}


\maketitle

\section{Introduction}

The magnetorotational instability (MRI) is a mechanism by which the
magnetic field can destabilize a hydrodynamically stable flow of a
conducting fluid without altering its velocity distribution. The MRI
was first discovered theoretically in cylindrical Taylor-Couette (TC)
flow of perfectly conducting fluid subject to an axial magnetic field
\cite{Velikhov-1959,Chandrasekhar-1960}. Three decades later, Balbus
and Hawley \cite{Balbus-Hawley-1991} suggested that the MRI may account
for the fast formation of stars by driving turbulent transport of
angular momentum in accretion disks. This proposition has triggered
not only numerous theoretical studies \cite{Balbus-Hawley-1998} but
also several attempts to reproduce the MRI in the laboratory \cite{Sisan-etal,Nature-2006}.
A major challenge to such experiments is posed by the parameter known
as the magnetic Reynolds number $\Rm$, which is required to be at
least $\sim10$ for the MRI to set in. For a typical liquid metal,
characterized by a small magnetic Prandtl number $\Pm\sim10^{-5}-10^{-6},$
this translates into a large hydrodynamic Reynolds number $\RE=\Rm/\Pm\sim10^{6}-10^{7}$
\cite{Goodman-Ji-2002}. At such large Reynolds numbers, the flow
on which the MRI is expected to develop may become turbulent due to
purely hydrodynamic instabilities \cite{Sisan-etal}. 

A way to circumvent this problem was proposed by Hollerbach and R\"udiger
\cite{HR-05}, who suggested that a magnetorotational-type instability
can take place in cylindrical TC flow at $\RE\sim10^{3}$ when the
imposed magnetic field is helical rather than purely axial as for
the standard MRI (SMRI). An instability resembling this new type helical
MRI (HMRI) was shortly thereafter observed in the PROMISE experiment
\cite{Rued-apjl,Stefani-etal,Stefani-NJP}. These observations were
disputed by Liu \emph{et al.} \cite{Liu-etal2006}, who found no such
instability in their inviscid theoretical analysis of finite length
cylinders with insulating end caps. Subsequently, the observed phenomenon
was conjectured to be a transient growth rather than a self-sustained
instability \cite{Liu-etal2007,Liu2008,Liu2009}. This proposition
was, in turn, opposed by Priede and Gerbeth \cite{Priede-Gerbeth2009},
who showed that there is not only a convective but also an absolute
HMRI threshold. Thus, a self-sustained HMRI can experimentally be
observed in a system of sufficiently large axial extension. However,
the comparison with the experimental results revealed that the HMRI
has been observed slightly beyond the range of its absolute instability,
where it is expected according to ideal TC flow model. This discrepancy
implies a deviation of the real base flow from the TC one, which may
be caused by the end effects. One such end effect disturbing the base
flow in the original PROMISE experiment was due to the copper end
cap, which allowed radial electric current to connect over the liquid
gap between the cylinders \cite{Priede2009}. This flaw was corrected
in the modified PROMISE experiment, where an insulating end cap was
used, which, in order to reduce the Ekman pumping, was split into
two separately rotating rings \cite{Stefani-etal2009a,Stefani-etal2009c}.
Although the instability appeared much sharper in the modified setup
than in the original one, its nature may still be questionable. Uncertainty
is due to the virtually unknown hydrodynamic stability limit of the
real base flow, whose deviation from ideal TC flow was significant
and practically unavoidable in the experiment. Without knowing the
actual hydrodynamic limit it is practically impossible to distinguish
the HMRI from a magnetically modified Taylor vortex flow \cite{Priede2009}. 

Regardless of the experimental reproducibility, the HMRI is also of
a questionable astrophysical relevance. Using a WKB analysis Liu \emph{et
al}. \cite{Liu-etal2006} showed that however the HMRI is able to
destabilize centrifugally stable velocity distributions, it does not
reach up to the astrophysically relevant Keplerian rotation profile.
This claim was doubted by R\"udiger and Hollerbach \cite{RH-07},
who pointed out that according to the numerical results for the TC
flow in a finite-width gap \cite{Priede-etal-2007}, the HMRI can
apparently reach the Keplerian rotation profile provided that at least
one of the boundaries is sufficiently conducting. However, it is important
to note that this conclusion concerns only the convective HMRI, which
is not generally self-sustained and requires an external excitation
to be effective. No destabilizing effect due to conducting boundaries
was observed for the absolute instability threshold \cite{Priede-Gerbeth2009}.
Although the extension of the absolute HMRI beyond the Rayleigh line
was found to increase with the strength and helicity of the magnetic
field, the previous study was unable to conclude whether the Keplerian
velocity profile is attainable. The problem was the excessive numerical
resolution required for the thin boundary layers developing in strong
magnetic field. However, the HMRI does not appear to be related with
the boundary layers, whose main function is to satisfy the no-slip
boundary condition imposed by the viscosity. These redundant boundary
layers can be eliminated by ignoring the viscosity, which appears
to be insignificant for astrophysical conditions. In this paper, fluid
is considered not only inviscid but also highly electrically resistive.
It means that the magnetic Prandtl number is assumed to be zero regardless
of the viscosity, which corresponds to the inductionless approximation
\cite{Priede-etal-2007}. 

The aim of the present work is to investigate numerically whether
the astrophysically relevant Keplerian rotation profile can be attained
by inviscid HMRI in cylindrical TC flow when a finite-width annulus
is considered. The obtained results show that only convective HMRI
mode can reach the Kepler limit provided that the boundaries are perfectly
conducting. However, this mode requires not only a permanent external
excitation to be observable but also has a long wave length, which
is incompatible with limited thickness of accretion disks. The absolute
HMRI as well as the convective one at insulating boundaries are found
to obey the Liu limit also for a finite-width annulus. This is because
the maximum extension of the HMRI beyond the Rayleigh line is attained
when the magnetic field is nearly azimuthal, which results in a short-wave
instability as in the local WKB approximation. 

The paper is organized as follows. In Sec. \ref{sec:prob} the problem
is formulated. The local WKB-type solution for a narrow gap is revisited
in Sec. \ref{sec:WKB}. Section \ref{sec:Num} presents numerical
results concerning the convective and absolute HMRI thresholds for
both insulating and perfectly conducting boundaries. The paper is
concluded with a summary of results in Sec. \ref{sec:end}.

\section{\label{sec:prob}Problem formulation}

\begin{figure}
\begin{centering}
\includegraphics[width=0.3\textwidth]{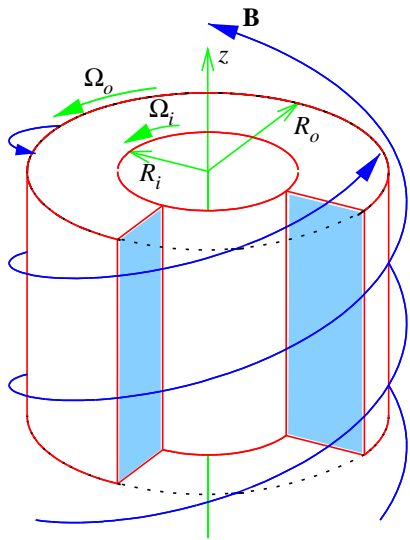} 
\par\end{centering}

\caption{(Color online) Sketch of the formulation of the problem.}
\end{figure}

Consider an incompressible inviscid fluid of electrical conductivity
$\sigma$ filling the gap between two infinite concentric cylinders
with the inner radius $R_{i}$ and the outer radius $R_{o}$ rotating
with the angular velocities $\Omega_{i}$ and $\Omega_{o}$, respectively,
in the presence of an externally imposed steady magnetic field $\vec{B}_{0}=B_{\phi}\vec{e}_{\phi}+B_{z}\vec{e}_{z}$
with axial and azimuthal components $B_{z}=B_{0}$ and $B_{\phi}=\beta B_{0}R_{i}/r$
in the cylindrical coordinates $(r,\phi,z),$ where $\beta$ is a
dimensionless parameter characterizing the geometrical helicity of
the magnetic field (see Fig. 1). The induced magnetic field is assumed
to be negligible relative to the imposed one, which corresponds to
the inductionless approximation. The velocity of inviscid fluid flow
$\vec{v}$ is governed by Euler equation with electromagnetic body
force \begin{equation}
\frac{\partial\vec{v}}{\partial t}+(\vec{v}\cdot\vec{\nabla})\vec{v}=\rho^{-1}\left(-\vec{\nabla}p+\vec{j}\times\vec{B}_{0}\right),\label{eq:N-S}\end{equation}
 where the induced current follows from Ohm's law for moving medium
\begin{equation}
\vec{j}=\sigma\left(\vec{E}+\vec{v}\times\vec{B}_{0}\right).\label{eq:Ohm}\end{equation}
 In addition, we assume the characteristic time of velocity variation
to be much longer than the magnetic diffusion time $\tau_{0}\gg\tau_{m}=\mu_{0}\sigma L^{2},$
which leads to the quasi-stationary approximation, according to which
$\vec{\nabla}\times\vec{E}=0$ and $\vec{E}=-\vec{\nabla}\Phi,$ where
$\Phi$ is the electrostatic potential \cite{Roberts-1967}. Mass
and charge conservation imply $\vec{\nabla}\cdot\vec{v}=\vec{\nabla}\cdot\vec{j}=0.$

As long as the viscosity is not exactly zero, the problem admits a
base state with a purely azimuthal velocity distribution $\vec{v}_{0}(r)=\vec{e}_{\phi}v_{0}(r),$
where the velocity profile \[
v_{0}(r)=r\frac{\Omega_{o}R_{o}^{2}-\Omega_{i}R_{i}^{2}}{R_{o}^{2}-R_{i}^{2}}+\frac{1}{r}\frac{\Omega_{o}-\Omega_{i}}{R_{o}^{-2}-R_{i}^{-2}}\]
is independent of the viscosity and thus holds also in the inviscid
limit. 

Note that the base flow is not affected by the magnetic field because
the latter gives rise only to the electrostatic potential $\Phi_{0}(r)=B_{0}\int v_{0}(r)dr,$
whose gradient compensates the induced electric field so that there
is no current in the base state $(\vec{j}_{0}=0)$. However, current
may be induced in the perturbed state \[
\left\{ \begin{array}{c}
\vec{v},p\\
\vec{j},\Phi\end{array}\right\} (\vec{r},t)=\left\{ \begin{array}{c}
\vec{v}_{0},p_{0}\\
\vec{j}_{0},\Phi_{0}\end{array}\right\} (r)+\left\{ \begin{array}{c}
\vec{v}_{1},p_{1}\\
\vec{j}_{1},\Phi_{1}\end{array}\right\} (\vec{r},t)\]
 where $\vec{v}_{1},$ $p_{1},$ $\vec{j}_{1},$ and $\Phi_{1}$ are
small-amplitude perturbations for which Eqs. (\ref{eq:N-S}) and (\ref{eq:Ohm})
after linearization take the form \begin{eqnarray}
\frac{\partial\vec{v}_{1}}{\partial t}+(\vec{v}_{1}\cdot\vec{\nabla})\vec{v}_{0}+(\vec{v}_{0}\cdot\vec{\nabla})\vec{v}_{1} & = & \rho^{-1}\left(-\vec{\nabla}p_{1}+\vec{j}_{1}\times\vec{B}_{0}\right)\nonumber \\
\vec{j}_{1} & = & \sigma\left(-\vec{\nabla}\Phi_{1}+\vec{v}_{1}\times\vec{B}_{0}\right).\label{eq:j1}\end{eqnarray}
 In the following, we focus on the axisymmetric perturbations which
are typically much more unstable than the nonaxisymmetric ones \cite{Rued-ANN}.
The nonaxisymmetric HMRI modes recently found by Hollerbach, Teeluck
and R\"udiger \cite{Hollerbach-etal-2010} in purely azimuthal magnetic
field lay outside the scope of this study. For axisymmetric perturbations,
the solenoidality constraints are satisfied by meridional stream functions
for fluid flow and electric current as \begin{eqnarray*}
\vec{v} & = & v\vec{e}_{\phi}+\vec{\nabla}\times\psi\vec{e}_{\phi},\\
\vec{j} & = & j\vec{e}_{\phi}+\vec{\nabla}\times h\vec{e}_{\phi}.\end{eqnarray*}
Note that $h$ is the azimuthal component of the induced magnetic
field, which is used subsequently instead of $\Phi$ for the description
of the induced current. Thus, we effectively retain the azimuthal
component of the induction equation to describe the meridional components
of the induced current, while the azimuthal current is explicitly
related to the radial velocity. The use of the electrostatic potential
$\Phi,$ which provides an alternative mathematical formulation for
the induced currents in the inductionless approximation, would result
in slightly more complicated governing equations. In addition, for
numerical purposes, we introduce an auxiliary variable, vorticity
\[
\vec{\omega}=\vec{\nabla}\times\vec{v}=\omega\vec{e}_{\phi}+\vec{\nabla}\times v\vec{e}_{\phi}.\]
Perturbation is sought in the normal mode form \begin{equation}
\left\{ v_{1},\omega_{1,}\psi_{1},h_{1}\right\} (\vec{r},t)=\left\{ \hat{v},\hat{\omega},\hat{\psi},\hat{h}\right\} (r)\times e^{\gamma t+ikz},\label{eq:pert}\end{equation}
 where $\gamma$ is, in general, a complex growth rate and $k$ is
the axial wave number, which is real for the conventional stability
analysis and complex when the absolute instability is considered.
Henceforth, we proceed to dimensionless variables by using $R_{i},$
$1/\Omega_{i},$ $R_{i}\Omega_{i},$ $B_{0},$ and $\sigma B_{0}R_{i}\Omega_{i}$
as the length, time, velocity, magnetic field, and current scales,
respectively. The nondimensionalized governing equations read as \begin{eqnarray}
\gamma\hat{v} & = & ikr^{-1}(r^{2}\Omega)'\hat{\psi}+\Lambda ik\hat{h},\label{eq:vhat}\\
\gamma\hat{\omega} & = & 2ik\Omega\hat{v}-\Lambda ik(ik\hat{\psi}+2\beta r^{-2}\hat{h}),\label{eq:omghat}\\
0 & = & D_{k}\hat{\psi}+\hat{\omega},\label{eq:psihat}\\
0 & = & D_{k}\hat{h}+ik(\hat{v}-2\beta r^{-2}\hat{\psi}),\label{eq:hhat}\end{eqnarray}
where $D_{k}f\equiv r^{-1}\left(rf'\right)'-(r^{-2}+k^{2})f$ and
the prime stands for $d/dr;$ $\Lambda=\sigma B_{0}^{2}/\rho\Omega_{i}$
is the Elsasser number, also referred to as the magnetohydrodynamic
interaction parameter,which will be the main dimensionless parameter
for the subsequent analysis;\begin{equation}
\Omega(r)=\frac{\lambda^{-2}-\mu+r^{-2}\left(\mu-1\right)}{\lambda^{-2}-1}\label{eq:Omeg}\end{equation}
is the dimensionless angular velocity of the base flow defined in
terms of $\lambda=R_{o}/R_{i}$ and $\mu=\Omega_{o}/\Omega_{i}.$
Note that Eqs. (\ref{eq:vhat})--(\ref{eq:hhat}) are invariant upon
the transformations $\{\Omega,k,\hat{h}\}\rightarrow-\{\Omega,k,\hat{h}\}$
and $\{\beta,k,\hat{v}\}\rightarrow-\{\beta,k,\hat{v}\},$ which means
that reversing the direction of rotation $\Omega$ or that of helicity
$\beta$ merely inverts the direction in which perturbations propagate
along the axis. Similarly, Eqs. (\ref{eq:vhat})--(\ref{eq:hhat})
are invariant upon the transformation $(k,\gamma)\rightarrow(-k,\gamma*)$
applied together with the complex conjugate operation, which implies
that for a fixed wave number the reversion of $\Omega$ or $\beta$
inverts only the sign of frequency $\omega=\Im[\gamma].$ Owing to
these symmetries it suffices to consider only positive $\beta$.

In inviscid approximation, the flow perturbation on the inner and
outer cylinders at $r=1$ and $r=\lambda,$ respectively, satisfy
the impermeability condition $\hat{\psi}=0.$ Boundary conditions
for $\hat{h}$ on insulating and perfectly conducting cylinders are
$\hat{h}=0$ and $(r\hat{h})'=0,$ respectively. There are no boundary
conditions imposed on the azimuthal velocity perturbation, which is
defined in terms of the two previous variables by Eq. (\ref{eq:vhat}).

Equations (\ref{eq:vhat})--(\ref{eq:hhat}) for perturbation amplitudes
were solved in the same way as in Refs. \cite{Priede-Gerbeth2009,Priede-etal-2007}
using a spectral collocation method based on the Chebyshev-Lobatto
grid with a typical number of internal points $N=32,$ which ensured
the accuracy of about five digits.

\section{\label{sec:WKB}Narrow-gap approximation}

Before undertaking numerical solution of the whole problem, it is
instructive to revisit the local WKB-type solution based on the narrow-gap
approximation defined by $\delta=\lambda-1\ll1,$ which is the dimensionless
gap width between the cylinders. Representing the cylindrical radius
as $r=1+\delta s,$ where $s$ is a local radial coordinate, the angular
base flow velocity (\ref{eq:Omeg}) in Eq. (\ref{eq:omghat-n}) at
the leading order in $\delta$ becomes $\Omega\approx1.$ The dimensionless
radial gradient term of the angular momentum in Eq. (\ref{eq:vhat})
can be written for arbitrary gap width as\begin{equation}
r^{-1}(r^{2}\Omega)'=2\frac{\lambda^{-2}-\mu}{\lambda^{-2}-1}=2(1+\Ro),\label{eq:M-rad}\end{equation}
which is constant not only for narrow but also finite-width gap. The
latter relation can be used to define the Rossby number for arbitrary
gap width as \begin{equation}
\Ro=\frac{1-\mu}{\lambda^{-2}-1},\label{eq:Ro}\end{equation}
which for a narrow gap reduces to $\Ro\approx(\mu-1)/(2\delta)\approx(\ln\sqrt{\Omega})'$.
Taking $\delta$ as the length scale, which implies the substitutions
$\{k,\beta\}\rightarrow\{k,\beta\}/\delta,$ $\{\hat{\psi},\hat{h}\}\rightarrow\{\hat{\psi},\hat{h}\}\delta^{2}$
and $\hat{v}\rightarrow\hat{v}\delta,$ Eqs. (\ref{eq:vhat})--(\ref{eq:hhat})
reduce at the leading order to\begin{eqnarray}
\gamma\hat{v} & = & 2ik(1+\Ro)\hat{\psi}+\Lambda ik\hat{h},\label{eq:vhat-n}\\
\gamma\hat{\omega} & = & 2ik\hat{v}-\Lambda ik(ik\hat{\psi}+2\beta r^{-2}\hat{h}),\label{eq:omghat-n}\\
0 & = & \hat{\psi}''-k^{2}\hat{\psi}+\hat{\omega},\label{eq:psihat-n}\\
0 & = & \hat{h}''-k^{2}\hat{h}+ik(\hat{v}-2\beta r^{-2}\hat{\psi}),\label{eq:hhat-n}\end{eqnarray}
where the prime now stands for $d/ds.$

A locally periodic solution can be sought in the form $\{\hat{v},\hat{\psi,}\hat{\omega,}\hat{h}\}(s)=\{\hat{v}_{0},\hat{\psi}_{0},\hat{\omega}_{0},\hat{h}_{0}\}\sin(ls)$
with constant amplitudes and the radial wave number $l.$ Substituting
this into Eqs. (\ref{eq:vhat-n})--(\ref{eq:hhat-n}) and eliminating
the amplitudes yields the dispersion relation \begin{equation}
4\left(ik(1+\Ro)+\beta\Lambda\kappa^{2}\right)(ik+\beta\Lambda\kappa^{2})-(\gamma+\Lambda\kappa^{2})\left((\gamma+\Lambda\kappa^{2})K^{2}+4\beta^{2}\Lambda\kappa^{2}\right)=0,\label{eq:disp}\end{equation}
where $K^{2}=k^{2}+l^{2}$ and $\kappa=k/K.$ 

Although the quadratic equation (\ref{eq:disp}) is easy solvable
for the growth rate $\gamma,$ it is instructive first to consider
asymptotic solutions for small and large interaction parameters$\Lambda.$
In the nonmagnetic case $\Lambda=0$, the solution is simple: \begin{equation}
\gamma_{0}=\pm2i\kappa\sqrt{1+\Ro},\label{eq:gamma0}\end{equation}
which describes exponentially growing Taylor vortices due to the centrifugal
instability when $\Ro<-1,$ and constant-amplitude inertial waves
when $\Ro\ge-1.$ For strong interaction at $\Lambda\gg1$, Eq. (\ref{eq:disp}),
reducing to\[
4(\beta\Lambda\kappa^{2})^{2}-(\gamma+\Lambda\kappa^{2})\left((\gamma+\Lambda\kappa^{2})K^{2}+4\beta^{2}\Lambda\kappa^{2}\right)=0,\]
yields \begin{equation}
\gamma=-\Lambda\kappa^{2}\left[1+2(\beta/K)^{2}\left(1\pm\sqrt{1+(\beta/K)^{-2}}\right)\right]<0,\label{eq:N>>1}\end{equation}
which shows that all perturbations are magnetically damped. For weak
interaction at $\Lambda\ll1,$ the growth rate may be sought as a
perturbation of the nonmagnetic solution $\gamma\approx\gamma_{0}+\Lambda\gamma_{1},$
where $\gamma_{0}$ is defined by Eq. (\ref{eq:gamma0}). This expression
substituted into Eq. (\ref{eq:disp}) yields\begin{equation}
\gamma_{1}=-K^{-2}\left[k^{2}+2(\beta\kappa)^{2}\pm\beta\kappa k\frac{2+\Ro}{\sqrt{1+\Ro}}\right].\label{eq:gamma1}\end{equation}
According to the above expression, there may be a finite range of
helicities bounded by \begin{equation}
\beta_{\pm}=\frac{K}{4}\left[\frac{2+\Ro}{\sqrt{1+\Ro}}\pm\sqrt{\frac{(2+\Ro)^{2}}{1+\Ro}-8}\right],\label{eq:bt-pm}\end{equation}
 in which inertial waves turn unstable $(\gamma_{1}>0)$ however small
the interaction parameter $\Lambda$. It means that the inviscid HMRI,
similarly to its standard counterpart, can be triggered by an arbitrary
weak magnetic field provided that its helicity lies in the range defined
above. For such an instability range to exist, $\beta_{\pm}$ has
to be real, which, in turn, requires the term under the square root
to be nonnegative and yields \begin{equation}
\pm\Ro\ge2\sqrt{2}\pm2=\pm\Ro_{L}^{\pm}.\label{eq:Ro-pm}\end{equation}
Taking the plus and minus signs in this expression yields $\Ro\geq2+2\sqrt{2}=\Ro_{L}^{+}$
and $\Ro\leq2-2\sqrt{2}=\Ro_{L}^{-},$ respectively, which are originally
due to Liu \emph{et al}. \cite{Liu-etal2006}. 

The damping of all perturbations at $\Lambda\gg1$ following from
Eq. (\ref{eq:N>>1}) implies that instability, if any, is possible
only at sufficiently small $\Lambda.$ The marginal interaction parameter
$\Lambda_{c},$ at which perturbations become neutrally stable, is
defined by $\Re[\gamma]=0.$ Substituting $\gamma=i\omega,$ where
$\omega$ is the frequency of neutrally stable perturbations, into
Eq. (\ref{eq:disp}) and taking the imaginary part, we obtain\[
\omega=\frac{2(2+\Ro)\kappa}{K/\beta+2\beta/K}.\]
This expression substituted into the real part of Eq. (\ref{eq:disp})
results in \begin{equation}
\Lambda_{c}=\sqrt{\omega^{2}/\kappa^{2}-4(1+\Ro)}=2\sqrt{\left(\frac{2+\Ro}{K/\beta+2\beta/K}\right)^{2}-(1+\Ro)}.\label{eq:Nc}\end{equation}
Again, for instability to exist, $\Lambda_{c}$ has to be real, which
requires the term under the square root to be non-negative. This is
equivalent to\[
\frac{(2+\Ro)^{2}}{1+\Ro}\ge(K/\beta+2\beta/K)^{2}=f(K/\beta)\ge f(\sqrt{2})=8,\]
which results in the same constraint as that for $\Lambda\ll1$ given
by Eq. (\ref{eq:Ro-pm}). When the condition above is satisfied, instability
is confined to the finite wave number band defined by the limiting
values, \begin{equation}
K_{\pm}=\frac{\beta}{2}\left(\frac{2+\Ro}{\sqrt{1+\Ro}}\pm\sqrt{\frac{(2+\Ro)^{2}}{1+\Ro}-8}\right)\sim\beta,\label{eq:K-pm}\end{equation}
at which $\Lambda_{c}$ in Eq. (\ref{eq:Nc}) turns to zero. 

These are the basic characteristics of the HMRI in the narrow-gap
approximation. It is important to note that although this instability
can affect some centrifugally stable velocity distributions with $\Ro>-1,$
it does not extend up to the astrophysically relevant Keplerian profile
$\Omega\sim r^{-3/2},$ whose Rossby number (\ref{eq:Ro}), \begin{equation}
\Ro_{K}=-\frac{3}{4}>\Ro_{L}^{-}\approx-0.828,\label{eq:Ro-K}\end{equation}
lies outside the Liu range (\ref{eq:Ro-pm}). The aim of the following
section is to investigate numerically whether this constraint can
be overcome when a finite-width gap is considered.

\section{\label{sec:Num}Numerical results for a finite-width gap}

The numerical results presented in this section are for the gap width
between the cylinders equal to the radius of the inner cylinder, which
corresponds to $\lambda=R_{o}/R_{i}=2.$ The corresponding Rayleigh
line is $\mu_{R}=\Omega_{o}/\Omega_{i}=\lambda^{-2}=0.25.$ Above
this critical ratio of rotation rates of the inner and outer cylinders
the specific angular momentum $r^{2}\Omega$ turns radially outward
increasing and thus the flow becomes hydrodynamically stable with
respect to axisymmetric inviscid perturbations.

\subsection{Convective instability}

\begin{figure}
\begin{centering}
\includegraphics[width=0.5\textwidth]{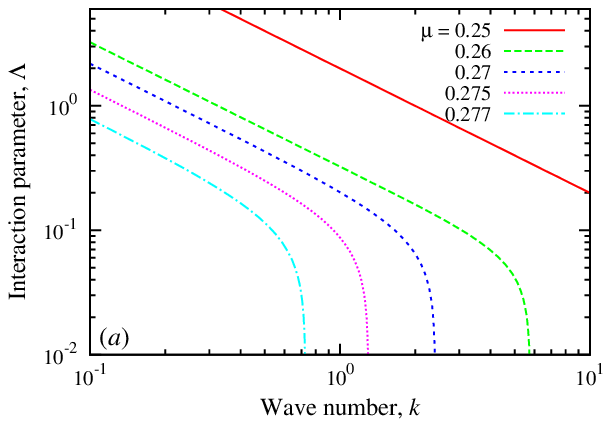}\includegraphics[width=0.5\textwidth]{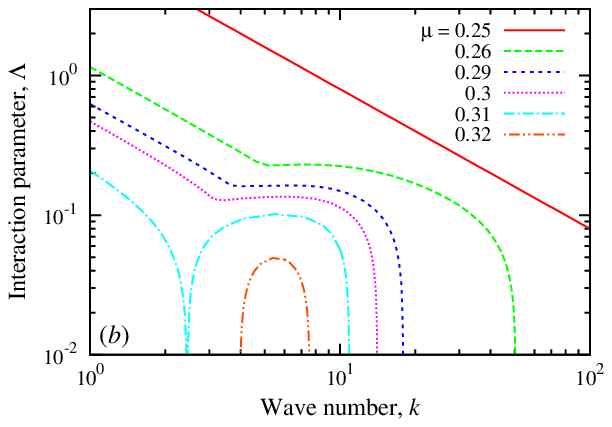} 
\par\end{centering}

\caption{\label{fig:Nk_bt-i}(Color online) The marginal interaction parameter
$\Lambda$ against the wave number $k$ for the helicities (a) $\beta=2$
and (b) $\beta=8$ at different $\mu;$ $\Lambda=\beta/k$ at the
Rayleigh line $\mu=0.25.$}
\end{figure}

In this section, real wave numbers are considered $(k_{i}=0)$ as
in the standard linear stability analysis, which corresponds to the
so-called convective instability. At this threshold, flow becomes
able to amplify certain perturbations, which are not necessarily self-sustained
and thus may require a permanent external excitation to be observable.
For self-sustained perturbations, the absolute instability is required,
which is considered in the next section.

\begin{figure}
\begin{centering}
\includegraphics[width=0.5\textwidth]{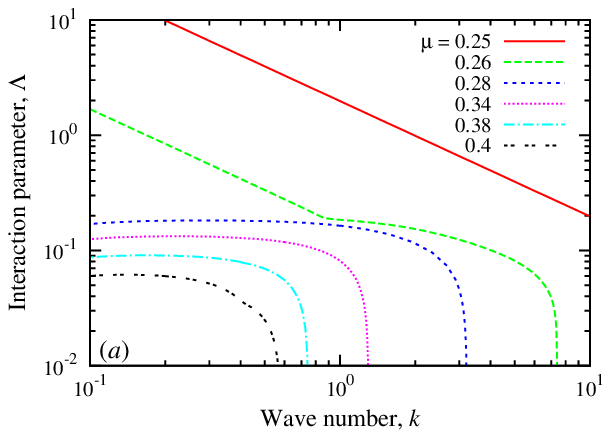}\includegraphics[width=0.5\textwidth]{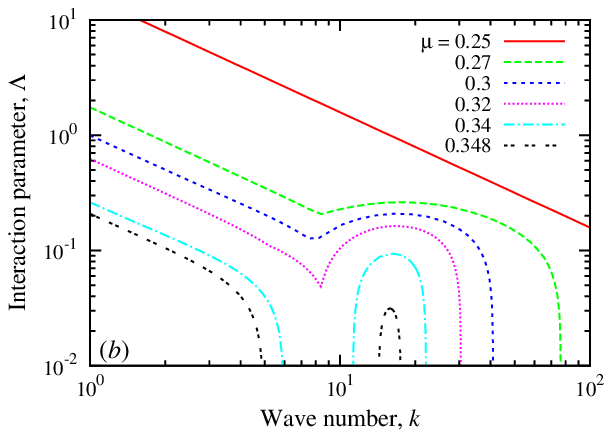} 
\par\end{centering}

\caption{\label{fig:Nk_bt-c}(Color online) The marginal interaction parameter
$\Lambda$ against the wave number $k$ for the helicities (a) $\beta=2$
and (b) $\beta=16$ at different $\mu$; $\Lambda=\beta/k$ at the
Rayleigh line $\mu=0.25.$}
\end{figure}

\begin{figure}
\begin{centering}
\includegraphics[width=0.5\textwidth]{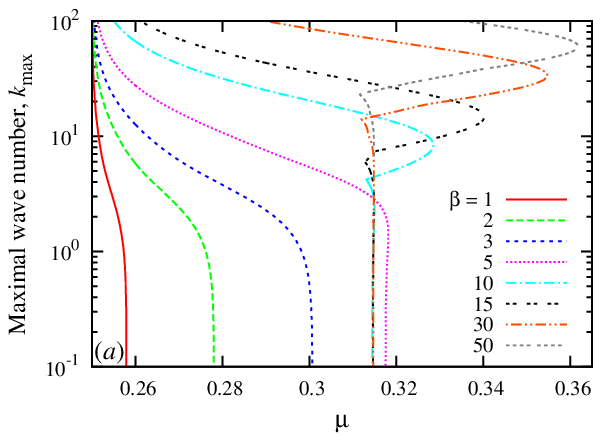}\includegraphics[width=0.5\textwidth]{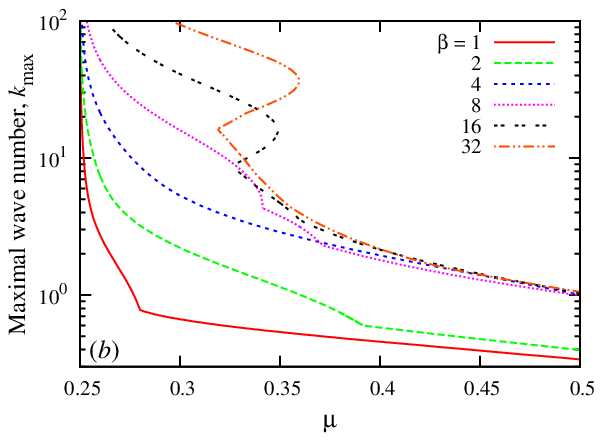} 
\par\end{centering}

\caption{\label{fig:wkmu_bt}(Color online) Extension of the HMRI beyond the
Rayleigh line $(\mu=0.25)$ for (a) insulating and (b) perfectly conducting
boundaries at various helicities $\beta.$ }
\end{figure}

The marginal interaction parameters below which the temporal growth
rate $\gamma_{r}$ becomes positive, are shown in Figs. \ref{fig:Nk_bt-i}
and \ref{fig:Nk_bt-c} against the wave number of the corresponding
mode for insulating and perfectly conducting boundaries, respectively.
For the centrifugally marginal state defined by $\mu=\mu_{R},$ the
neutral stability boundary turns out to be particularly simple: $\Lambda=\beta/k.$
The origin of this simple relation is not obvious. Slightly beyond
the Rayleigh line $(\mu>0.25),$ the short-wave unstable modes disappear.
With the increase in $\mu,$ the range of unstable modes rapidly shrinks
toward the longer waves. As seen in Fig. \ref{fig:Nk_bt-i}(a), for
sufficiently low helicities $\beta,$ there is a critical $\mu$ at
which the long-wave instability mode disappears altogether. For higher
$\beta,$ the instability disappears in a more complex way. As seen
in Fig. \ref{fig:Nk_bt-i}(b) for $\beta=8$, there is an intermediate-wave
mode, which outlasts the long-wave one. This longer lasting mode affects
only a limited range of wave numbers, which also quickly shrinks from
both ends with the increase in $\mu.$ As a result, the range of unstable
wave numbers disappears altogether by shrinking to a point at some
critical $\mu,$ which slightly exceeds that for the disappearance
of the long-wave mode discussed above. As seen in Fig. \ref{fig:wkmu_bt},
the critical wave number at which this mode disappears increases with
the helicity $\beta.$ Thus, this intermediate-wave mode resembles
the narrow-gap instability considered in the previous section. 

For perfectly conducting boundaries and sufficiently low helicities,
the marginal interaction parameter, which is shown in Fig. \ref{fig:Nk_bt-c}(a)
against the wave number for $\beta=2$, demonstrates a similar variation
to that for insulating boundaries seen in Fig. \ref{fig:Nk_bt-i}(a).
In contrast to the insulating boundaries, with the increase in $\mu$,
the original long-wave instability mode quickly disappears. What is
left behind resembles an intermediate-wave mode for insulating boundaries
at high helicities except that the remaining mode extends up to the
long-wave limit $k=0.$ In contrast to the original long-wave mode
at $\mu=\mu_{R},$ the marginal interaction parameter for this mode
tends to a finite value rather than growing as $\sim k^{-1}$ for
$k\rightarrow0.$ The main difference from the insulating case is
that instability does not vanish at a finite $\mu.$ With the increase
in $\mu,$ instability just shifts to longer waves with the marginal
interaction parameter approaching zero, which, as shown in the previous
section, implies a similar decrease also in the growth rate. The variation
of the marginal interaction parameter for perfectly conducting boundaries
at higher helicities is shown in Fig. \ref{fig:Nk_bt-c}(b) for $\beta=16.$
Slightly beyond the Rayleigh line, instability is due to an intermediate-wave
mode, which vanishes at a finite $\mu$ similarly to that for the
insulating boundaries seen in Fig. \ref{fig:Nk_bt-i}(b). What is
left behind is a long-wave mode, which persists at large $\mu$ by
shifting to ever longer waves as in the case of low helicities considered
above. The above results are summarized in Fig. \ref{fig:wkmu_bt},
which shows the extension of instability beyond the Rayleigh line
in terms of the marginal wave numbers at which the instability vanishes
plotted against the ratio of rotation rates $\mu$ for various helicities
$\beta.$

\subsection{\label{sub:abso}Absolute instability}

The convective instability considered in the previous section is not
necessarily self-sustained. In spatially unbounded systems, small
amplitude perturbations become self-sustained above the absolute instability
threshold. In addition to the positive temporal growth rate supposed
by the convective instability, the absolute instability requires zero
group velocity for the critical perturbation. This additional constraint
can be satisfied by a nonzero imaginary part of the wave number $k_{i}\not=0.$
From a physical as well as computational point of view, it is advantageous
to consider the absolute instability as a limiting case of the global
instability. The latter supposes that two oppositely propagating modes
with the same frequency can be mutually coupled by the reflections
from remote end walls. Two such marginal modes having the same imaginary
but, generally, different real parts of the wave number can form a
neutrally stable global mode with a finite wave packet length fitting
in a system of sufficiently large extension \cite{Lifshitz-Pitaevskii-81,Priede-Gerbeth-97}.
The global instability turns into the absolute one when two such oppositely
propagating modes merge so that also the real parts of the wave number
coincide, which results in a wave packet of infinite extension. This
approach will be pursued in the following to locate the absolute instability.

\begin{figure}
\begin{centering}
\includegraphics[width=0.5\textwidth]{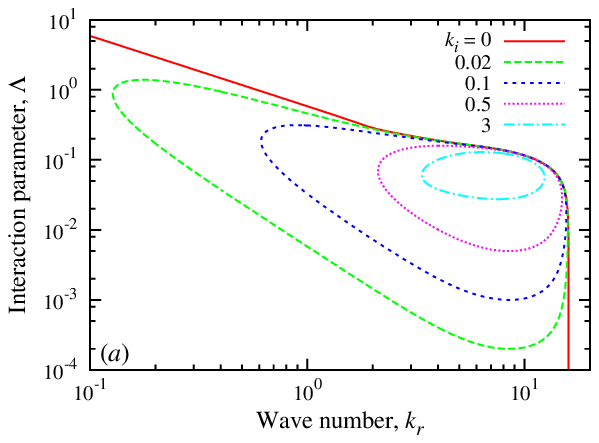}\includegraphics[width=0.5\textwidth]{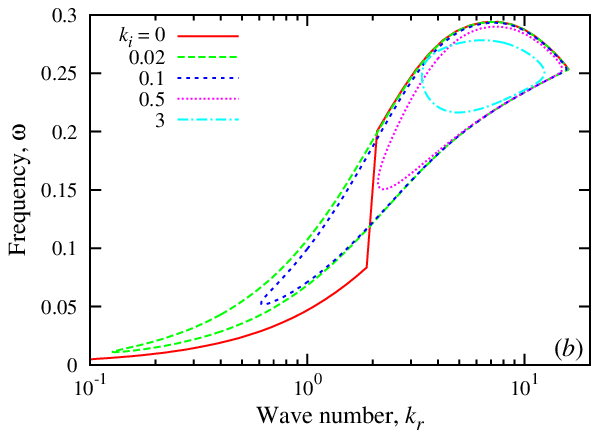} 
\par\end{centering}

\caption{\label{fig:Nki-kr}(Color online) Interaction parameter $\Lambda$
(a) and the frequency $\omega$ (b) for marginally stable modes $(\gamma_{r}=0)$
versus the real part of the wave number $k_{r}$ for $\mu=0.27$ and
$\beta=5$ at various imaginary parts of the wave number $k_{i}.$}
\end{figure}

\begin{figure}
\begin{centering}
\includegraphics[width=0.5\textwidth]{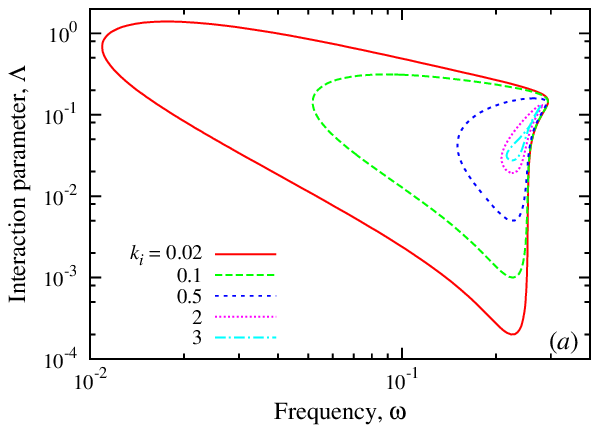}\includegraphics[width=0.5\textwidth]{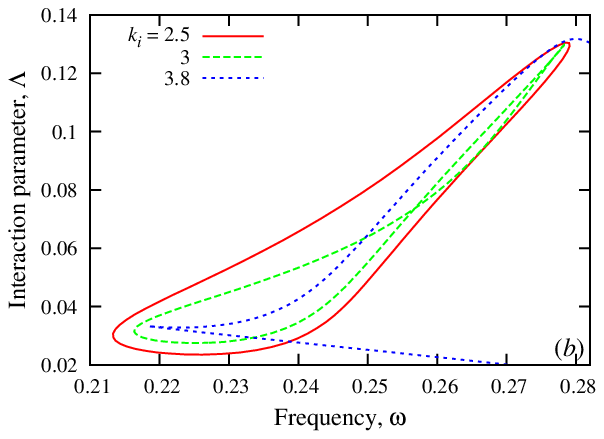}
\par\end{centering}

\caption{\label{fig:Nki-frq}(Color online) The marginal interaction parameter
$\Lambda$ versus the frequency of neutrally stable modes $\omega$
for $\mu=0.7$ and $\beta=5$ at various imaginary parts of the wave
number $k_{i}.$}
\end{figure}

The first step is to consider the neutral stability curves for $k_{i}\not=0$
which are plotted in Fig. \ref{fig:Nki-kr} for insulating boundaries
with a rotation rate ratio slightly beyond the Rayleigh line $(\mu=0.27)$
and a moderate helicity $\beta=5.$ As discussed in the previous section,
the neutral stability curve for $k_{i}=0,$ which defines the convective
instability threshold, consists of an intermediate- and a long-wave
branch. Transition between these branches shows up as a break on the
neutral stability curve and as a jump in the associated frequency,
which are seen in Figs. \ref{fig:Nki-kr}(a) and \ref{fig:Nki-kr}(b),
respectively. For $k_{i}>0$ representing the main interest here,
both branches merge together and neutral stability curves start to
form closed loops, which tighten as $k_{i}$ is increased. A more
relevant information follows from Fig. \ref{fig:Nki-kr}, where the
previous two quantities are plotted against each other. It is important
to notice that the curves in Fig. \ref{fig:Nki-frq}(b) form not only
closed loops but also intersect themselves in a certain range of $k_{i}.$
For a fixed $k_{i},$ the frequencies and the interaction parameters
of two modes coincide at the intersection. As discussed above, two
such waves could sustain each other by the reflections from the end
walls and thus form a marginal global mode provided that they propagate
in the opposite directions. The direction of propagation can be determined
from the variation of the interaction parameter with $k_{i}.$ Namely,
two modes propagate in the opposite directions if the interaction
parameter of one intersecting branch increases while the other decreases
with a small variation in $k_{i}$ \cite{Priede-Gerbeth-97}. Since
this is the case seen in Fig. \ref{fig:Nki-frq}(b), the intersection
point defines a marginal interaction parameter for a neutrally stable
global mode consisting of two oppositely traveling waves with different
real parts of the wave number. At $k_{i}=3.80,$ the loop below the
intersection tightens together forming a cusp. This critical point
at which both modes merge together represents the threshold of the
absolute instability. Besides the lower threshold there is also an
upper one, which occurs at $k_{i}=3.28$ when another cusp forms at
the top of the loop. This upper cusp is formed by the second intersection
which is hardly noticeable at the upper tip of the loop for $k_{i}=3$
in Fig. \ref{fig:Nki-frq}(b). 

\begin{figure}
\begin{centering}
\includegraphics[width=0.5\textwidth]{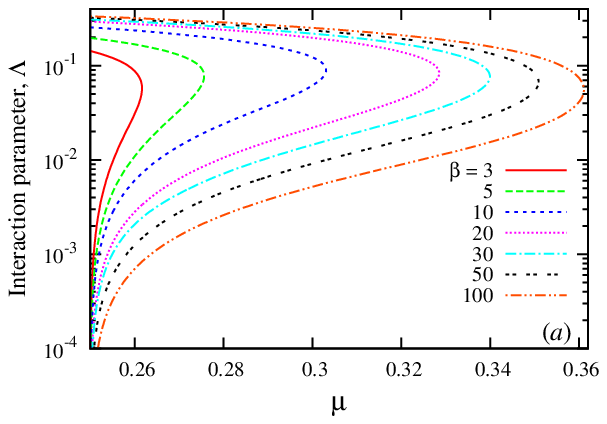}\includegraphics[width=0.5\textwidth]{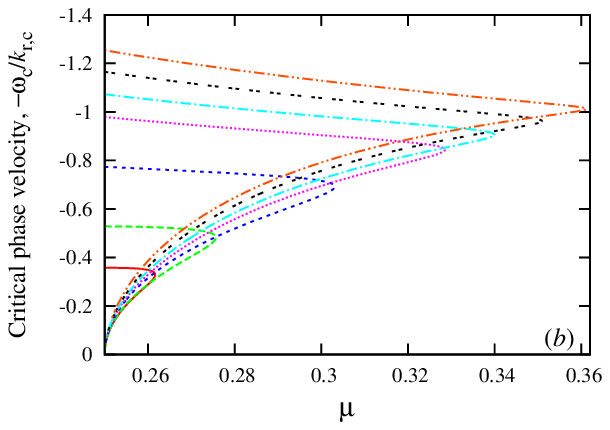}
\par\end{centering}

\begin{centering}
\includegraphics[width=0.5\textwidth]{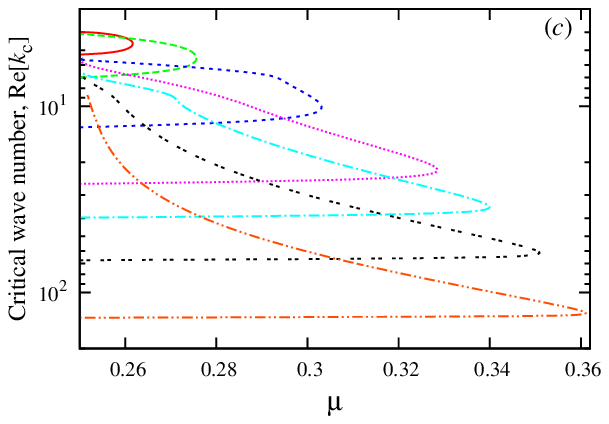}\includegraphics[width=0.5\textwidth]{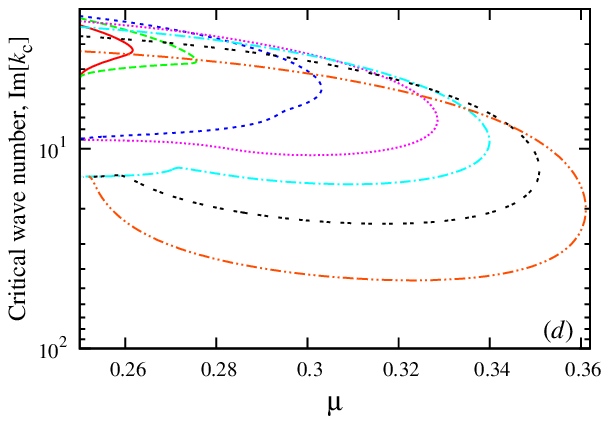}
\par\end{centering}

\caption{\label{fig:Ng-mu}(Color online) The critical interaction parameter
$\Lambda_{c}$ (a), the phase velocity (b), the real (c) and imaginary
(d) parts of the critical wave number of the absolute instability
versus $\mu$ for insulating boundaries at various helicities $\beta.$ }
\end{figure}

The thresholds of the absolute instability for insulating boundaries
at various helicities are plotted in Fig. \ref{fig:Ng-mu} against
the ratio of rotation rates $\mu.$ With increasing the rotation rate
or reducing the strength of the magnetic field, the HMRI sets in below
the upper critical value of the interaction parameter and vanishes
below the lower one, which are shown in Fig. \ref{fig:Ng-mu}(a).
Such a double threshold is a characteristic feature of the HMRI that
distinguishes it from a magnetically modified Taylor vortex flow,
which exists below the Rayleigh line and has only the upper threshold.
The critical phase velocity and the real and imaginary parts of wave
number at the instability thresholds are shown in Figs. \ref{fig:Ng-mu}(b)--\ref{fig:Ng-mu}(d),
respectively. Although the increase of the helicity $\beta$ makes
the instability extend farther beyond the Rayleigh line, the extension
is seen in Fig. \ref{fig:mu-bt} to saturate at large $\beta.$ 

\begin{figure}
\begin{centering}
\includegraphics[width=0.5\textwidth]{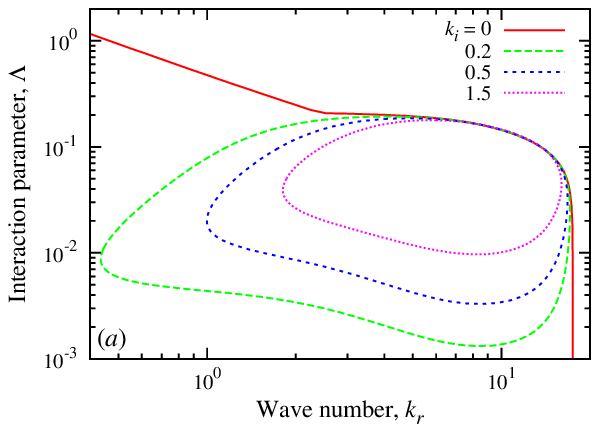}\includegraphics[width=0.5\textwidth]{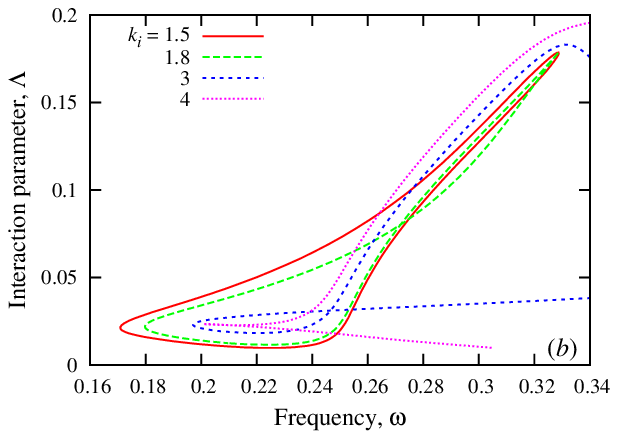} 
\par\end{centering}

\caption{\label{fig:N-kic}(Color online) The marginal interaction parameter
$\Lambda$ against the wave number $k_{r}$ (a) and against the frequency
$\omega$ (b) for perfectly conducting boundaries at $\mu=0.27,$
$\beta=5$ and various $k_{i}.$}
\end{figure}

\begin{figure}
\begin{centering}
\includegraphics[width=0.5\textwidth]{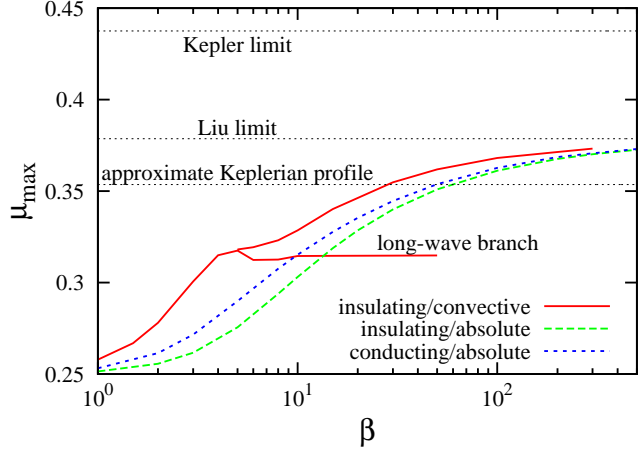} 
\par\end{centering}

\caption{\label{fig:mu-bt}(Color online) The extension of the convective and
absolute HMRI beyond the Rayleigh line versus the helicity $\beta$
for insulating and perfectly conducting boundaries.}
\end{figure}

For the absolute instability, in contrast to the convective one considered
in the previous section, there is no significant difference between
insulating and perfectly conducting boundaries. For perfectly conducting
boundaries, only the intermediate but not the long-wave modes can
turn absolutely unstable. As seen in Fig. \ref{fig:N-kic}(a), the
neutrally stable modes with $k_{i}>0$ are limited to finite wave
number and interaction parameter ranges. The marginal interaction
parameter plotted in Fig. \ref{fig:N-kic}(b) against the frequency
of neutrally stable modes forms self-intersecting loops as in the
case of insulating boundaries shown in Fig. \ref{fig:Nki-frq}(b).
Thus, the threshold of the absolute instability for perfectly conducting
boundaries turns out to be close to that for insulating boundaries
shown in Fig. \ref{fig:Ng-mu}. 

The proximity of the absolute instability thresholds for insulating
and perfectly conducting boundaries is demonstrated in Fig. \ref{fig:mu-bt},
which shows the extension of the HMRI beyond the Rayleigh line depending
on the helicity $\beta.$ Note that the extension of the convective
instability for perfectly conducting boundaries is not shown because
it appears unlimited in Fig. \ref{fig:wkmu_bt}(b). At high $\beta,$
the extension of the convective instability for insulating boundaries
as well as that of the absolute instability for both insulating and
perfectly conducting cylinders are seen to tend to a certain limit
of $\mu.$ The main question is whether this limit includes the Keplerian
velocity profile $\Omega\sim r^{-3/2}.$ A simple approximation of
the Keplerian profile with the TC one leads to $\mu_{\Delta}=\lambda^{-3/2}\approx0.35$
for $\lambda=2$ under consideration here. As seen in Fig. \ref{fig:mu-bt},
the HMRI certainly persists beyond this limit. On the other hand,
the lower Liu limit (\ref{eq:Ro-pm}) substituted into Eq. (\ref{eq:Ro})
yields \begin{equation}
\mu_{L}=1+(1-\lambda^{-2})\Ro_{L}^{-}=\frac{5}{2}-\frac{3}{\sqrt{2}},\label{eq:mu-L}\end{equation}
which appears to be the asymptote approached by $\mu$ in Fig. \ref{fig:mu-bt}.
It means that the HMRI obeys the Liu limit not only for a thin but
also for a finite-width gap. But this obviously contradicts the above
conjecture that the HMRI can affect the Keplerian profile, whose Rossby
number according to Eq. (\ref{eq:Ro-K}) lies beyond the Liu limit
and yields\begin{equation}
\mu_{K}=1+(1-\lambda^{-2})\Ro_{K}=\frac{7}{16}.\label{eq:mu-K}\end{equation}
Indeed, this ratio of rotation rates is seen in Fig. \ref{fig:mu-bt}
to lie significantly above that for the Liu limit. It is important
to note that $\mu_{\Delta}$ is an integral criterion based on the
difference of angular velocity over the gap width. The velocity difference
for the TC profile coinciding with that for the Keplerian profile
means that only the average velocity gradients of both profiles are
the same. Since the velocity profiles are different, the TC profile
will be in some parts of the gap shallower and in some other parts
steeper than that of the Keplerian profile \cite{RH-07}. Therefore,
simple approximation of the Keplerian profile by the TC one based
on the total velocity variation may be misleading. This becomes obvious
by considering the radial gradient term of the specific angular momentum
given by Eq. (\ref{eq:M-rad}), which determines the centrifugal stability.
This is also the key parameter for the HMRI, whose critical value
is modified by helical magnetic field as shown by the narrow-gap approximation.
Expression (\ref{eq:M-rad}) defines the local slope of the specific
angular momentum profile, which is constant for TC flow and can be
expressed in terms of either $\mu$ or $\Ro.$ Then it is easy to
see that the slope of TC flow profile corresponding to $\mu_{\Delta}$
is indeed considerably lower than the local slope of the Keplerian
profile defined by Eq. (\ref{eq:mu-K}). 

\section{\label{sec:end}Summary and conclusions}

This paper presented the analysis of the HMRI in the astrophysically
relevant limit of a vanishing viscosity. In order to focus on the
HMRI, the analysis was carried out in the inductionless approximation
defined by zero magnetic Prandtl number $\Pm=0$, which excludes the
standard MRI. These assumptions resulted in the problem defined by
a single dimensionless parameter, the Elsasser number, also referred
to as the MHD interaction parameter, which characterizes the ratio
of electromagnetic and inertial forces. Using a Chebyshev collocation
method convective and absolute instability thresholds were numerically
calculated with respect to axisymmetric perturbations of cylindrical
TC flow subject to a helical magnetic field. Without the magnetic
field, the flow becomes centrifugally unstable when the specific angular
momentum of fluid at the inner cylinder exceeds that at the outer
one. According to the previous WKB analysis, a strong magnetic field
has a stabilizing effect, but a weak field of certain helicity can
destabilize some centrifugally stable velocity distributions lying
beyond the so-called Rayleigh line. The latter defines the limit of
the centrifugal instability in terms of the critical ratio of rotation
rates for a given gap width between the cylinders. However, in the
WKB approximation, the extension of the instability beyond the Rayleigh
line is constrained by the so-called Liu limit and does not reach
up to the astrophysically relevant Keplerian velocity profile. The
obtained numerical results show that the Liu limit is obeyed also
by the nonlocal solution for a finite width gap between the cylinders
except for perfectly conducting boundaries. Maximal extension of the
absolute HMRI as well as that of the convective HMRI at insulating
boundaries is attained when the magnetic field is nearly azimuthal,
which produces a short-wave instability as in the WKB approximation.
The short-wave character of the instability explains why the nonlocal
solution obeys the Liu limit and why it may also apply to other gap
widths rather than just the case of $\lambda=2$ considered in this
study. Moreover, the short wave length implies that the absolute instability,
which formally applies to the systems of infinite axial extension,
may be relevant also for relatively thin disks. This is because the
absolute instability, as discussed at the beginning of Sec. \ref{sub:abso},
represents the limit of the global instability with respect to the
wave packets of finite length. The wave packets formed by two short
waves with slightly different wave numbers can fit in the relatively
thin disks. 

In the case of perfectly conducting boundaries, the instability can
extend sufficiently far beyond the Rayleigh line to reach the astrophysically
relevant Keplerian velocity profile. However, this is the case only
for the convective instability mode, which is not self-sustained and
thus requires an external excitation to be effective. Moreover, as
the ratio of rotation rates increases further beyond the Rayleigh
line, this instability mode shifts to ever longer waves. First, due
to its long wave length this mode is not only inherently nonlocal
but also unable to fit within limited thickness of accretion disks.
Second, the critical interaction parameter for this mode approaching
zero implies that its growth rate approaches zero too. As a result,
this weak instability can be suppressed by a finite viscous damping,
which also limits its extension beyond the Rayleigh line \cite{Priede-etal-2007,Priede-Gerbeth2009}.
Given these three basic constraints, the HMRI at perfectly conduction
boundaries seems of a little astrophysical relevance, too. 

Note that the Liu limit appears only in the inductionless approximation,
which, however, captures the essence of HMRI. At nonzero $\Pm,$ instability
can extend beyond the Liu limit \cite{HR-05}. Obviously, the HMRI
can extend similarly to the SMRI up to the Velikhov-Chandrasekhar
limit corresponding to the solid-body rotation. However, this extension
is due to the SMRI, in which HMRI merges with at a sufficiently high
Reynolds number \cite{Kirillov-Stefani2010}. For small $\Pm,$ the
merging point, at which the HMRI loses its identity by turning into
the SMRI, is located only slightly above the Liu limit \cite{Kirillov-Stefani2011}.
Thus, the HMRI cannot affect the Keplerian velocity profiles even
at finite $\Pm.$

\end{document}